# Code Reverse Engineering problem for Identification Codes

Julien Bringer[1] and Hervé Chabanne[1,2]
[1] Morpho, France.
[2] Télécom ParisTech, France.

**Abstract**

At ITW'10, Bringer *et al.* suggest to strengthen their previous identification protocol by extending the Code Reverse Engineering (CRE) problem to identification codes. We first extend security results by Tillich *et al.* on this very problem. We then prove the security of this protocol using information theoretical arguments.

**Keywords.** CRE, Identification Codes.

## 1 Introduction

At Indocrypt'09, Bringer *et al.* [3] introduce a new identification protocol based on the use of identification codes [1]. Their proposal, denoted here the BCCK identification protocol, relies on a construction of identification codes by Moulin and Koetter [12] using Reed-Solomon codes. In a few words, the BCCK identification protocol can be described as followed (cf. Figure 1, in [3] a low-cost contactless device (CLD) and its reader want to mutualy authenticate themselves).

The CLD stores two secret polynomials $P, P'$ of degree less than $k$ known only by the Reader; to authenticate itself to the CLD, the reader proves the knowledge of $P$ by sending $\langle i, P(\alpha_i)\rangle$ where $\alpha_i$ is the $i$-th element of $\mathbb{F}_q$. The CLD proves its identity by replying with $\langle P'(\alpha_i)\rangle$.

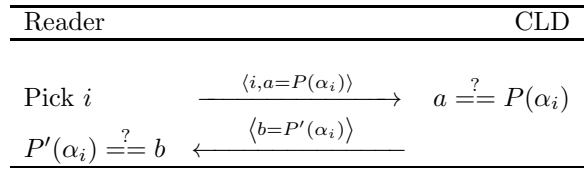

Figure 1: BCCK Identification Protocol [3]

[3] proves the security and the privacy properties of their protocol relying on

a classical cryptographic assumption known as the Polynomial Reconstruction (PR) problem [7–10].

We here want to switch from this computational perspective to an information theoretical one.

On one hand, a first attempt was made in this direction by [11] where some recommendations are made to reach this more stringent goal. On the other hand, [2] introduces an extension of the BCCK identification protocol, simply considering that the underlying Reed-Solomon codes stay unknown from the adversary who has thus to solve the Code Reverse Engineering (CRE) problem [4–6, 13] to recover the initial parameters of the BCCK identification protocol. This restriction on what is available to the adversary does not modify the underlying structure of the BCCK identification protocol.

In this paper, we show that, in fact, an adversary cannot solve this CRE for identification codes. Our results are based on those of [6]. We consider different cases taking into account the noise over the channel and the capacity of the adversary to isolate or not the communications of a CLD.

## 2 Related Works

### 2.1 Code Reverse Engineering Problem

The Code Reverse Engineering (CRE) problem [4–6, 13] corresponds to the situation where an observer tries to retrieve information from an eavesdropped communication without any specific prior knowledge on the encoding representation of the transmitted data.

In the CRE problem, it is assumed that the adversary knows the length $n$ of the encoded messages and a subset of codes of length $n$. Then by eavesdropping several messages over a noisy channel, he tries to determine from which code they are generated.

**Definition 1 (Code Reverse Engineering problem [6])** *Let $\mathcal{C}$ be a family of codes of given length $n$ and given rate $R$.*

- *Let $C$ be a code chosen randomly in $\mathcal{C}$ and $x^1, \ldots, x^M$ be $M$ random codewords to be transmitted over the communication channel.*

- *Given the received words $y^1, \ldots, y^M$, the problem is to guess which $C$ has been used.*

The difficulty depends on the number of received words, the level of noise and the rate of the codes. In [6], the authors analyze the number of eavesdropped messages that is needed to achieve a correct guess with good probability.

Let $V$ and $W$ be two random variables. Let $H(V)$, respectively $H(V|W)$ and $I(V;W)$, denote the binary entropy of $V$, respectively the conditional entropy of $V$ given $W$, and the mutual information of $V$ and $W$. When assuming that the codewords are chosen independently and that the channel is memoryless, we have the following result.

**Lemma 1 ( [6])** *The conditional entropy $H(C|\bar{y})$ of $C$ given $\bar{y} = (y^1, \ldots, y^M)$ is lower bounded by*

$$\log_2(\#\mathcal{C}) - M(I(x;y) - I((x;y)|C))$$

*with $x$ one of the $x^i$ and $y$ the corresponding $y^i$.*

This implies that the closer the mutual information $I((x|C);(y|C))$ and $I(x;y)$ are, the harder to guess $C$ is.

Thanks to Fano's inequality, [6] estimates also how large $M$ should be to achieve a good guess of $C$ with negligible error probability when $n$ and $\#\mathcal{C}$ go to infinity; this needs $M \simeq \frac{\log_2(\#\mathcal{C})}{(I(x;y) - I((x;y)|C))}$. The result is then exploited for linear codes with rate below the channel capacity (i.e. with overwhelming probability to decode correctly a received word when $n$ large) and for regular LDPC codes.

## 2.2 Identification Codes

Let $\mathcal{X}, \mathcal{Y}$ be two alphabets, $\eta$ the message length, and $W^\eta$ a channel from $\mathcal{X}^\eta$ to $\mathcal{Y}^\eta$, defined as a conditional probability law: $W^\eta(y|x)$ is the probability to receive a message $y \in \mathcal{Y}^\eta$ given a transmitted message $x \in \mathcal{X}^\eta$. By extension, for a given subset $E \subset \mathcal{Y}^\eta$, $W^\eta(E|x)$ is the probability to receive a message belonging to $E$ when $x$ has been transmitted.

**Definition 2 (Identification Code, [1])** *A $(\eta, N, \lambda_1, \lambda_2)$-identification code from $\mathcal{X}$ to $\mathcal{Y}$ is given by a family $\{(Q(\cdot|i), D_i)\}_i$ with $i \in \{1, \ldots, N\}$ where:*

- $Q(\cdot|i)$ *is a probability distribution over $\mathcal{X}^\eta$, that encodes $i$ (the encoding set of $i$ is defined as the set of messages $x$ for which $Q(x|i) > 0$, in other words, the set of messages likely to encode $i$),*

- $D_i \subset \mathcal{Y}^\eta$ *is the decoding set,*

- $\lambda_1$ *and $\lambda_2$ are the first-kind and second-kind error rates, with*

$$\lambda_1 \geq \sum_{x \in \mathcal{X}^\eta} Q(x|i) W^\eta(\overline{D_i}|x)$$

*and*

$$\lambda_2 \geq \sum_{x \in \mathcal{X}^\eta} Q(x|j) W^\eta(D_i|x)$$

*(where $W^\eta(D_i|x)$ is the probability to be in the decoding set $D_i$ given a transmitted message $x$ and $W^\eta(\overline{D_i}|x)$ the probability to be outside the decoding set)*

*for all $i, j \in \{1, \ldots, N\}$ such that $i \neq j$.*

To simplify the discussion in next sections, we restrict ourselves to the case where error rates inequalities above are in fact equalities (i.e. the error rates are the same for all choice of $i$ and $j$).

Moulin and Koetter introduce in [12] the following identification code based on Reed-Solomon codes.

A Reed-Solomon code over a finite field $\mathbb{F}_q$, of length $n < q-1$, and dimension $k$, is the set of the evaluations of all polynomials $P \in \mathbb{F}_q[X]$ of degree less than $k-1$, over a subset $F \subset \mathbb{F}_q$ of size $n$ ($F = \{\alpha_1, \ldots, \alpha_n\}$). In other words, for each $k$-tuple $(x_0, \ldots, x_{k-1}) \in \mathbb{F}_q^k$, the corresponding Reed-Solomon word is the $n$-tuple $(y_1, \ldots, y_n)$ where $y_i = \sum_{j=0}^{k-1} x_j \alpha_i^j$. In the sequel, we identify a source word $(x_0, \ldots, x_{k-1}) \in \mathbb{F}_q^k$ with the corresponding polynomial $P = \sum_{j=0}^{k-1} x_j X^j \in \mathbb{F}_q[X]$.

**Definition 3 (Moulin-Koetter RS-Identification Codes)** *Let $\mathbb{F}_q$ be a finite field of size $q$, $k \leq n \leq q-1$ and an evaluation domain $F = \{\alpha_1, \ldots, \alpha_n\} \in \mathbb{F}_q$.*

*Consider the collection $A_{F,P} = \{(i, P(\alpha_i)) | i \in \{1, \ldots, n\}\}$ for $P$ any polynomial on $\mathbb{F}_q$ of degree at most $k-1$.*

*Then the Moulin-Koetter RS-Identification Codes are defined by:*

- *their encoding distribution $Q(\cdot|i)$ which is taken as the uniform distribution over $A_{F,P}$,*

- *their family of encoding and decoding sets $\{(A_{F,P}, A_{F,P})\}_{P \in \mathbb{F}_q[X], \deg P < k}$.*

From the definition and the fact that the Reed-Solomon codes are Maximum Distance Separable, this leads to ($\eta = \log_2 n + \log_2 q$, $N = q^k$, $\lambda_1 = 0$, $\lambda_2 = \frac{k-1}{n}$) identification codes from $\{0,1\}$ to $\{0,1\}$.

**Example 1** *Throughout this paper, we take the parameters suggested in [3] for the identification protocol of Figure 1: $q = 2^{64}$, $n = 2^{11}$, $k = 2^8$.*

*In the original paper, there is a limitation on the number of times a CLD can be identified by a reader. [3] preconises that a same CLD can only be interrogated at most 2048 times.*

## 3 CRE for identification codes

In this section, we show how to extend the Code Reverse Engineering problem and the bounds from Section 2.1 to the case of identification codes. This is the first important contribution of this paper.

Definition 1 in the context of transmission codes gives the following definition for identification codes.

**Definition 4 (Identification CRE problem)** *Let $\mathcal{C}$ be a family of identification codes (cf. Definition 2) of given parameters $(\eta, N, \lambda_1, \lambda_2)$ from the alphabet $\mathcal{X}$ to alphabet $\mathcal{Y}$.*

- Let $C = \{(Q(\cdot|i), D_i)\}_{i \in \{1,\ldots,N\}}$ be a code chosen randomly in $\mathcal{C}$ and $\bar{i} = (i^1, \ldots, i^M)$ be $M$ random messages to be encoded over the channel.

- Given the received messages $\bar{x} = (x^1, \ldots, x^M)$, the problem is to guess which $C$ has been used.

**Remark 1** *Note that we modify the original problem, replacing the encoded messages by the messages to be encoded, to be able to address the case without errors in the BCCK identification protocol.*

With a memoryless channel, Lemma 1 is adapted accordingly as follows.

**Lemma 2** *For independent choices of $\bar{i} = (i^1, \ldots, i^M)$, the conditional entropy $H(C|\bar{x})$ of the identification code $C$ given the received messages $\bar{x} = (x^1, \ldots, x^M)$ is lower bounded by*

$$\log_2(\#\mathcal{C}) - M(I(i;x) + I((x;C)|i) - I((i;x)|C)) \tag{1}$$

*i.e.*

$$\log_2(\#\mathcal{C}) - M(H(x) - H(i) + H(i|C,x) - H(x|C,i)) \tag{2}$$

*Proof.* As for the original proof of Lemma 1, this is based on the relation $I(i;x;C) = I(i;x) + I((x;C)|i) = I(x;C) + I((i;x)|C)$ where here $I((x;C)|i)$ is not equal to 0. As $H(i) = H(i|C)$, this leads to $\log_2(\#\mathcal{C}) - M(I((x;C)|i) - I((i;C)|x))$. $I((x;C)|i) - I((i;C)|x)$ can also be simplified into $H(x) - H(i) + H(i|C,x) - H(x|C,i)$ as $H(x|i) - H(i|x) = H(x) - H(i)$. □

One important difference with the CRE problem for transmission codes is that the solution is not trivial for a noiseless channel. This is due to the first-kind $\lambda_1$ and second-kind $\lambda_2$ error rates of the identification codes: if at least one error-rate is non zero, then $I(i;x) < H(i)$ whereas the mutual information between a message and the received message would have been maximal with a transmission code. Intuitively, what makes the problem harder for identification codes is that the quantity of transmitted information can be very low.

Assume that the distribution is regular in the error rate inequalities of Definition 2 (the same probability $W$ holds for all $x$):

- The first-kind error rate $\lambda_1$ implies that for a given $i$, we have a probability $\lambda_1$ to take a $x$ outside the decoding set of $i$. This means that $H(i|x, C)$ would be almost equal to $H(i)$. Consequently, with probability $\lambda_1$ we have $I(i;x) \approx I((i;x)|C)$.

- The second-kind error rate means that for another given message $j$ we have a probability $\lambda_2$ to also have $x \in D_j$. This implies that $H(i|x, C)$ may remain high.

**Corollary 1** *Let $\mathcal{X} = \mathcal{Y} = \{0,1\}$. For a constant size of the encoding*

$$\#\{x \mid Q(x|i) > 0\} = \tau(N)2^\eta$$

*in the context of a noiseless channel. The equation Eq.(2) simplifies itself into*

$$\log_2(\#\mathcal{C}) - M(\log_2 1/\tau(N)) \qquad (3)$$

*Proof.* The equation Eq.(2) becomes

$$\log_2(\#\mathcal{C}) - M(H(x) - \log_2 N + H(i|C,x) - \log_2 \tau(N)2^\eta)$$

and we have $H(x) \leq \eta$, $H(i) = \log_2 N$, $H(x|C,i) = \log_2 \tau(N)2^\eta$, and $\log_2 N \geq H(i|x,C)$. □

We see that the greater $\tau(N)$ will be, the greater the expression (3) will be. By using Fano's inequality as in [6], we obtain that we need $M \simeq \frac{\log_2(\#\mathcal{C})}{\log_2 1/\tau(N)}$ for guessing $C$ with negligible error probability when $n$ and $\#\mathcal{C}$ go to infinity. $M$ will go to infinity quickly as soon as $\log_2 1/\tau(N)$ is negligible compared to $\log_2(\#\mathcal{C})$.

In case of additional noise on the communication channel, the difficulty will increase with the level of noise (as for the classical CRE problem).

## 4 Application to BCCK protocol

Now comes the main contribution of our work: we use the CRE problem for identification codes to study the security of the BCCK identification protocol from an information theory perspective.

[11] suggests to increase the security of the identification protocol [3], which relies on the Polynomial Reconstruction problem, by additionally exploiting the Code Reverse Engineering (CRE) problem. The goal is to restrict further the information available to an eavesdropper or an active adversary.

Let $\mathbb{F}_q$ be a finite field of size $q$, $k \leq n \leq q-1$, we define $\mathcal{C}$ the set of the Moulin-Koetter ($\eta = \log_2 n + \log_2 q, N = q^k, \lambda_1 = 0, \lambda_2 = \frac{k-1}{n}$) identification codes. Following Definition 3, an identification code $C \in \mathcal{C}$ is defined according some evaluation domain $F_C = \{\alpha_{C,1}, \ldots, \alpha_{C,n}\} \in \mathbb{F}_q$ with

- a family of encoding and decoding sets $\{(A_{F_C,P}, A_{F_C,P})\}_{P \in \mathbb{F}_q[X], \deg P < k}$,
- where $A_{C,P} = \{(j, P(\alpha_{C,j})) | j \in \{1, \ldots, n\}\}$.

This doing, a random code $C$ is determined by the random choice of $n$ elements in $\mathbb{F}_q$. The size of $\mathcal{C}$ is $\binom{q}{n}n!$.

Lemma 2 and the same analysis that for Eq. (3) with $\tau(N) = 1/q$ lead to the following result.

**Corollary 2** $H(C|x^1, \ldots, x^M) \geq \log_2 \binom{q}{n}n! - M \log_2 q$

This underlines the difficulty of the CRE problem in this setting when $n$ grows to infinity for $M$ polynomial in $\log_2 n$: via Stirling's formula, $\log_2 \binom{q}{n}n!$ is approximately $q \log_2(q/(q-n)) + n \log_2(q-n) = \Omega(n)$.

**Example 2** *Taking back the values from Example 1,
this leads to $(75, 2^{64 \times 256}, 0, 1/8)$-identification codes. This gives:*

$$H(C|x^1, \ldots, x^M) \geq \log_2 \binom{2^{64}}{2^{11}} 2^{11}! - 64 \times M.$$

*As $\log_2 \binom{2^{64}}{2^{11}} 2^{11}!$ is approximately equal to $2^{17}$, taking $M = 2048$ makes the lower bound useless.*

*Consequently, note that we need a more stringent result than the one we just obtained. This motivates the introduction of a new tighter lower bound.*

With the following result, which is specific to the Moulin-Koetter construction, we prove that an adversary gains no information in this situation.

**Proposition 1** *For the $\mathcal{C}$ family of Moulin-Koetter identification codes defined as above over $\mathbb{F}_q$ with $k \leq n \leq q - 1$, for independent choices of $M$ messages $P^1, \ldots, P^M$ to be encoded for a random choice of $C \in \mathcal{C}$, we have*

$$H(C|x^1, \ldots, x^M) = \log_2 \binom{q}{n} n!$$

$x^1, \ldots, x^M$ *are the received messages, independently and randomly chosen in the encoding sets of $P^1, \ldots, P^M$, and eavesdropped by the adversary (here without noise).*

*Proof.* We have $H(x) = \log_2 n + \log_2 q$, $H(P) = \log_2 N = k \log_2 q$, $H(x|C, P) = \log_2 n$ and we know that the uncertainty on $P$ knowing $x$ and $C$ corresponds to the choice of a polynomial of degree at most $k-2$, i.e. $H(P|x, C) = (k-1) \log_2 q$. □

### 4.1 Validity of the independence assumption

In the protocol of [3], the messages to be encoded and the encoding messages are not fully independent due:

1. to the relative small size of the encoding sets and the correlation between them (one can detect if the same index $j$ is used to encode independent polynomials),

2. to the potential ability of an adversary to detect whether two encoding messages are sent to the same CLD (i.e. that they are related to the same polynomial/message to be encoded).

We study now the impact on the previous estimation.
A more general version of Lemma 2 is given below for the situation where independent choices are not required.

**Lemma 3** Let $C \in \mathcal{C}$ be an identification code. Let $\bar{i} = (i^1, \ldots, i^M)$ be the variable associated to $M$ messages to be encoded and $\bar{x}$ the variable for the corresponding received messages. We have

$$H(C|\bar{x}) \geq \log_2(\#\mathcal{C}) \tag{4}$$
$$-(H(\bar{x}) - H(\bar{i}) + H(\bar{i}|C,\bar{x}) - H(\bar{x}|C,\bar{i}))$$

As $H(\bar{i})$ is always greater than $H(\bar{i}|C,\bar{x})$, we obtain

$$H(C|\bar{x}) \geq \log_2(\#\mathcal{C}) - (H(\bar{x}) - H(\bar{x}|C,\bar{i}))$$

Moreover, $H(\bar{x}) - H(\bar{x}|C,\bar{i}) \leq H(\bar{x}) \leq M \times H(x)$, thus

$$H(C|\bar{x}) \geq \log_2(\#\mathcal{C}) - M \times H(x) \tag{5}$$

With the parameters of the identification protocol, we deduce

$$H(C|\bar{x}) \geq \log_2 \binom{q}{n} n! - M(\log_2 n + \log_2 q)$$

This means that the knowledge of the adversary, on the code $C \in \mathcal{C}$ that is used, is still negligible when $n$ grows to infinity for $M$ polynomial in $\log_2 n$.

**Example 3** For the parameters from Example 2, $q = 2^{64}$, $n = 2^{11}$, $k = 2^8$, the lower bound is

$$\log_2 \binom{2^{64}}{2^{11}} 2^{11}! - 75 \times M$$

which remains high only for $M < 1747$.

We now study a tighter estimation directly from the expression (4). To correspond to the situation where the adversary is able to determine whether the same CLD is aimed (for instance by capturing it), assume that in $\bar{i} = (i^1, \ldots, i^M)$, each message $i = P$ is repeated $l$ times exactly. We assume that each time a different encoded message $x = (j, P(\alpha_j))$ is used. We obtain for $l \leq k$, the same result as before (cf. Proposition 1), whereas the entropy decreases for $l > k$.

**Proposition 2** For $l \leq k$, $H(C|\bar{x}) = \log_2 \binom{q}{n} n!$
For $l > k$, $H(C|\bar{x}) \geq \log_2 \binom{q}{n} n! - \frac{M}{l}(l-k) \log_2 q$

*Proof.* We apply Equation (4) with the following values:

- $H(\bar{x}) = \frac{M}{l} \sum_{j=0}^{l-1}(\log_2(n-j) + \log_2 q)$
- $H(\bar{i}) = \frac{M}{l} \log_2 N = \frac{M \times k}{l} \log_2 q$
- $H(\bar{i}|C,\bar{x}) = \frac{M}{l} \times (k-l) \log_2 q$ if $l \leq k$, 0 otherwise.
- $H(\bar{x}|C,\bar{i}) = \frac{M}{l} \sum_{j=0}^{l-1}(\log_2(n-j))$

□

Following the intuition, with non independent messages, the adversary gains some knowledge on the chosen code $C \in \mathcal{C}$ only when the order of repetition is strictly greater than $k$.

**Example 4** *We set the same parameters as in Example 2, $q = 2^{64}$, $n = 2^{11}$, $k = 2^8$. The lower bound is*

$$\log_2 \binom{2^{64}}{2^{11}} 2^{11}! - 64 \times \frac{M}{l}(l - 256)$$

*For instance, this remains high until about $M \leq 2^{19}$ with $l = 257$ and $M \leq 2^{12}$ for $l = 512$. The number of needed $M$ to have a low lower bound decreases up to approximately 2341 while increasing $l$ to $n$.*

*On the one hand, this can be interpreted as follows. A passive eavesdropping of 2048 BCCK identifications of the same CLD may enable an adversary to get almost all information on the underlying code when $M = 2341$. On the other hand, note that with only 2340 such eavesdroppings, the entropy stays very high for the adversary.*

# Acknowledgement

The authors thank Jean-Pierre Tillich for encouraging them to carrefully read [6]. They also thank Gérard Cohen for his comments.